\documentstyle [12pt] {article}

\begin{document}

\baselineskip = 20 true pt

\title{\huge \bf Growth of surfaces generated by a probabilistic
cellular automaton}

\author{~\\ ~\\ {\LARGE Pratip Bhattacharyya}
                        \thanks{E-mail : pratip@cmp.saha.ernet.in}\\ ~\\
 \large Low Temperature Physics Division\\
         Saha Institute of Nuclear Physics\\
         Sector - 1, Block - AF, Bidhannagar\\
         Calcutta 700 064, India.}

\date{}

\maketitle

\vspace{1.5cm}

\begin{abstract}

\noindent A one-dimensional cellular automaton with a probabilistic
evolution rule can generate stochastic surface growth in $(1 + 1)$
dimensions. Two such discrete models of surface growth are
constructed from a probabilistic cellular automaton which is
known to show a transition from a active phase to a absorbing
phase at a critical probability associated with two particular
components of the evolution rule. In one of these models,
called model $A$ in this paper, the surface growth is defined
in terms of the evolving front of the cellular automaton on
the space-time plane. In the other model, called model $B$,
surface growth takes place by a solid-on-solid deposition
process controlled by the cellular automaton configurations
that appear in successive time-steps. Both the models show
a depinning transition at the critical point of the generating
cellular automaton. In addition, model $B$ shows a kinetic
roughening transition at this point. The characteristics
of the surface width in these models are derived by scaling
arguments from the critical properties of the generating cellular
automaton and by Monte Carlo simulations.

\end{abstract}

\newpage

\section{Introduction}

\indent Cellular automata are discrete mathematical systems that
follow non-equilibrium dynamics~\cite{Wolfram1983}. Typically
a discrete variable with a finite number of possible states
evolves in discrete time-steps, simultaneously at all points
of a discrete space (for example, a lattice) on which it has
been defined, following a local evolution rule. The simplest
of these forms, known as elementary cellular automata,
have two states per site and only nearest-neighbour
interactions. The evolution of a $d$-dimensional cellular
automaton is by itself a process of surface growth in
$(d+1)$-dimensions. Besides, it can generate other surface
growth models, for example, by considering the configurations
of a cellular automaton in successive time-steps as sieves
with pores at the sites that have a particular state and
through these pores particles are deposited on the
corresponding sites of a substrate lattice.

\indent A cellular automaton with a probabilistic evolution rule
will generate a stochastic growth process. When the noise due
to the probability is of a non-thermal kind it may bring about
a phase transition even in a one-dimensional system with
short range interactions~\cite{Grassberger1984}. Such a phase
transition in a cellular automaton will induce a corresponding
effect in the surface growth models generated from it.
Generally two kinds of phase transitions are observed in
surface growth models : (1) morphological transitions such
as kinetic roughening transitions, and (2) depinning transitions.
While morphological transitions are due to a critical change in
the rate of growth or deposition, depinning transitions occur
by a critical increase in the force that drives the surface
so that it overcomes the pinning force \cite{Barabasi1995}.
Examples of such transitions in $(1 + 1)$-dimensions are few,
most of which are related to a Directed Percolation process :
kinetic roughening transitions in $(1 + 1)$-dimensions have been
observed in models of polynuclear growth~\cite{Kertesz1989}, growth
by absorption and desorption of solid on solid~\cite{Alon1996-98}
and fungal growth~\cite{Lopez1998} whereas depinning
transitions in $(1 + 1)$-dimensions were originally observed
in models of quenched point disorder~\cite{Tang1992} and porous
media~\cite{Buldyrev1992}.

\indent This paper reports a study of surface growth phenomena
generated by a one-dimensional probabilistic cellular automaton
that shows a phase transition in the universality class of
Directed Percolation.

\section{Two models with depinning transition}

\indent The one-dimensional probabilistic cellular automaton (PCA),
which is the generator of two surface growth models studied in this
paper, is defined as a line of sites with a binary variable
$a_i \in \{ 0, 1 \}$ assigned to each site $i$. A site is said to be
occupied if $a_i = 1$ and unoccupied otherwise. The automaton evolves
from any given initial configuration $\{ a^{(0)}_i \}$ by updating
the variables at all lattice sites simultaneously at each time-step.
Each site follows a local rule of evolution that involves only
nearest-neighbour interactions. The rule is specified by a set of
eight components $[a^{(t)}_{i-1}, a^{(t)}_i, a^{(t)}_{i+1}] \mapsto
a^{(t+1)}_i$ corresponding to the $2^3$ distinct three-site
neighbourhoods of interaction :

\begin{equation}
{t : \over {t+1 :}}~~~~~~{111 \over 0}~~~~{110 \over 0}~~~~{011 \over 0}~~~~{101 \over 0}~~~~{010 \over 1}~~\underbrace{{100 \over  }~~~~{001 \over  }}_{\begin{array}{lll}
                                                                                                                                                                  1 & \mbox{with probability} & p\\
                                                                                                                                                                  0 & \mbox{with probability} & 1 - p
                                                                                                                                                           \end{array}}~~{000 \over 0}
\label{eq:model}
\end{equation}

\noindent The probabilistic behaviour enters the cellular automaton
as a non-thermal noise added specifically to two mutually symmetric
rule components.

\vspace{0.5cm}

\indent Based on the evolution of this PCA from disordered
(uncorrelated) initial states two different models of surface
growth in $(1 + 1)$ dimensions are constructed :

\vspace{0.5cm}

\noindent (1)~Model $A~-$ The initial configuration of the PCA
forms the substrate with the occupied lattice sites acting as
growth centers. The propagating front of occupied sites
on the space-time plane defines a surface. Here growth occurs
in the direction of increasing time which therefore is the vertical
direction along which the surface height will be measured.
Overhangs may exist on the front but they are ignored so that
the height profile is a single-valued function. The height of
the surface $h_i (t)$ above a point $i$ on the substrate at time
$t$ is given by the time coordinate of the highest occupied site
in the $i$-th column which means that in this model all sites
in a column in the space-time plane between the substrate site
and the highest occupied site are assumed to be filled.
This is a growth model of the propagating interface type.

\vspace{0.5cm}

\noindent (2)~Model $B~-$ This is a model of the deposition type.
Surface growth takes place on a one-dimensional substrate (a line
of empty sites) by depositing particles in successive time-steps
according to the configuration of the cellular automaton at those
times. At time $t$ a particle is added to the $i$-th column, i.e.,
the height of the $i$-th column is increased by unity if
the corresponding site in the PCA is occupied at that time
($a^{(t)}_i = 1$). The height variable is defined by :

\begin{equation}
h_i (t) = \sum_{\tau = 1}^t a^{(\tau)}_i \, .
\label{eq:heightB}
\end{equation}

\noindent A surface grown in this way does not have overhangs.
Figure 1 shows typical examples of surfaces of models $A$
and $B$ generated by the evolving cellular automaton.
To illustrate the procedure of growth in the two models outlined
above the detailed mechanics are shown schematically for a small
part of the lattice.

\indent The purpose of having disordered initial states of the
PCA is to avoid the appearance of absorbing states at the beginning
of its evolution. The PCA is known to have three absorbing
states~\cite{Bhattacharyya1998} : the zero configuration
($\cdots 0000 \cdots$) and two mutually symmetric configurations
given by sequences of alternating $0$ and $1$ ($\cdots 0101 \cdots$,
$\cdots 1010 \cdots$). Occurence of these absorbing states
leads to either no growth (for the zero configuration) or
a trivial growth process (for the other two absorbing states).
A disordered state is constructed by assigning each lattice
site the value $1$ with a predefined probability and $0$ otherwise,
independently of the values at other lattice sites; when this
probability is $\frac{1}{2}$ the state is said to be completely
random.   In general, any initial state other than the three
absorbing states will serve this purpose.

\indent The growth models are characterised by the properties
of their surface width which, for a substrate of linear size
$L$, is defined in the usual way~\cite{Barabasi1995} :

\begin{equation}
w^2(L, t) = \left \langle \; \overline{\left [ h_i(t) -
 \bar{h}(t) \right ]^2} \; \right \rangle \, ,
 \label{eq:width}
\end{equation}

\noindent $\bar{h}(t) = \frac{1}{L} \sum_{i = 1}^L h_i(t)$ is the
average height of the surface above the substrate at time $t$. The bar
$\overline{~\cdots~}$ denotes average over the system and the angular
brackets $\langle \cdots \rangle$ denote configurational average.
In the addition to the width, a moving surface is characterised
by its average velocity of propagation, defined by~\cite{Barabasi1995} :

\begin{equation}
v_t = {\partial \over \partial t} \bar{h}(t) \, .
 \label{eq:velocity}
\end{equation}

\indent For the PCA model there exists a critical value of $p$,
observed to be $p_c \approx 0.75$~\cite{Bhattacharyya1996},
at which a continuous phase transition occurs with respect to
the density of occupied sites in the steady state; Monte Carlo
simulations~\cite{Bhattacharyya1998} of the PCA indicate that
the phase transition belongs to the universality class of
Directed Percolation. In the absorbing phase $(p < p_c)$
all initial configurations other than the two mutually
symmetric absorbing states following the rule in equation
(\ref{eq:model}) evolve to the zero configuration
(all sites unoccupied) which is a absorbing state of the PCA.
Evolution of the cellular automaton stops when the zero configuration
is reached. The presence of occupied sites in the cellular automaton
configuration at any time-step produces a moving surface in the
growth models. The appearance of the zero configuration results
in pinning of the surface \footnote{Though pinning centers and forces
are not explicit in these models, pinning of a surface growing in
$(d + 1)$-dimensions can be thought of as a $d$-dimensional
system entering an absorbing state where there are no particles.}.
In the active phase $(p > p_c)$ all initial configurations
other than the three absorbing states evolve to a active
steady state with a non-zero density of occupied sites.
This results in unpinned surfaces in the growth models.

\indent Thus the critical point $p_c$ of the cellular automaton
marks a depinning transition in the growth models, from a pinned
phase for $p < p_c$ to a moving phase for $p > p_c$. The order
parameter for the phase transition in the cellular automaton is
the density of occupied sites $\rho_\infty$ in the steady state
which in the supercritical region decreases continuously to zero
as the critical point is approached~\cite{Bhattacharyya1996} :

\begin{equation}
\rho_\infty \propto (p - p_c)^\beta,~~~~~~~p \to p_{c} \! ^{+} \, .
\label{eq:PCA-order}
\end{equation}

\noindent The appropriate order parameter in the growth models in
general is the asymptotic velocity $v_\infty$ of the surface in the
moving phase~\cite{Barabasi1995} :

\begin{equation}
v_\infty \propto (p - p_c)^{\theta},~~~~~~~p \to p_{c} \! ^{+} \, .
\label{eq:grow-order}
\end{equation}

\vspace{0.5cm}

\indent The following sections report the investigations
on the depinning transition in models $A$ and $B$ by scaling
arguments and Monte-Carlo simulations.

\section{Properties of Model $A$}

\indent By definition Model $A$ is the direct evolution of the PCA
from disordered initial states viewed as a growth process
that excludes surface overhangs. Therefore the properties
of the surface near the depinning transition can be
derived from those of the critical properties of the PCA.
Figure 2(a,b) shows typical examples of evolution of the surface width
$w(L, t)$ with time for substrates of different sizes $L$
obtained by Monte Carlo simulation of Model $A$. As it is usual for
correlated growth on substrates of finite size the surface width
gets saturated after a crossover time $t_{\times}$. Both above and
below the depinning transition point $p_c$ the saturated value of
the width, denoted by $w(L, \infty)$, shows two regimes separated
by a crossover length $L_\times$ $-$ (1) for $L \ll L_{\times}$ it
increases with $L$ indicating the appearance of a
\lq rough\rq~(or, $L$-dependent) regime; (2) for $L \gg L_{\times}$,
$w(L, \infty)$ tends to saturate to a constant value as
$L \to \infty$ which indicates a \lq smooth\rq~(or,
$L$-independent) regime. The values of $L_{\times}$ and
$w(\infty, \infty)$ depend only on the value of the noise
parameter $p$.  Therefore the saturated width $w(L, \infty)$ is
expected to follow a scaling relation of the form

\begin{equation}
{w(L, \infty) \over w(\infty, \infty)} \propto
         f_{<>} \left ({L \over L_{\times}} \right ) \, ,
 \label{eq:scale1}
\end{equation}

\noindent where $f_<$, $f_>$ denotes the scaling function in the
pinned $(p < p_c)$ and moving $(p > p_c)$ phases respectively.

\indent The crossover length $L_{\times}$ provides a characteristic
lengthscale of the model. Because of the scaling hypothesis that
the model has a unique diverging lengthscale at the critical point,
$L_{\times}$ must diverge as the correlation length of the PCA as
$p$ approaches $p_c$ from above and below :

\begin{equation}
L_{\times} \propto |p - p_c|^{- \nu_\perp} \, ,
 \label{eq:crosslength}
\end{equation}

\noindent where $\nu_\perp$ is the critical exponent for the correlation
length of the PCA. Since the growth of the surface occurs in the
direction of time, the saturated surface width is a measure of
temporal RMS fluctuations in the steady state. Consequently, in the
limit of an infinite substrate, the quantity $w(\infty, \infty)$
will diverge as the correlation time of the PCA :

\begin{equation}
w(\infty, \infty) \propto |p - p_c|^{- \nu_\parallel} \, ,
 \label{eq:width-Linfty-tinfty}
\end{equation}

\noindent This reduces the scaling law (\ref{eq:scale1}) to the
standard form of finite size scaling~\cite{Barber1983} :

\begin{equation}
w(L, \infty) \, |p - p_c|^{\nu_\parallel} \propto
         f_{<>}(L \, |p - p_c|^{\nu_\perp}) \, .
 \label{eq:scale2}
\end{equation}

\noindent Since the PCA is believed to be in the universality class of
Directed Percolation~\cite{Bhattacharyya1998}, the critical exponents
for $L_{\times}$ and $w(\infty, \infty)$ are expected to be
$\nu_\perp = \nu^{DP}_\perp \approx 1.097$ and $\nu_\parallel =
\nu^{DP}_\parallel \approx 1.734$ respectively~\cite{Jensen1996}.

\indent At $p_c$ the crossover length $L_{\times}$ diverges and
only the rough regime exists. The surface width is then expected
to follow the Family-Vicsek scaling law~\cite{Family1985} :

\begin{equation}
w(L, t) \propto L^{\alpha} f \left ({t \over L^z} \right ),
         ~~~~~~~~~p = p_c \, .
 \label{eq:scale3}
\end{equation}

\noindent The growth exponent $\lambda$ and the roughness exponent $\alpha$
are defined as $w(L, t) \propto t^{\lambda}$, $t \ll t_{\times}$
and $w(L, t) \propto L^{\alpha}$, $t \gg t_{\times}$ respectively
and the dynamic exponent $z$ occuring in (\ref{eq:scale3}) is given
by $t_{\times} \propto L^z$, $z = \alpha / \lambda$.
The value of the exponents $\alpha$, $\lambda$ and $z$ at $p_c$ can
be obtained by using the scaling hypothesis : corresponding to a
uniquely diverging lengthscale of the model there exists a uniquely
diverging timescale. The saturated width $w(L, \infty)$ in this model
measures temporal RMS fluctuations in the steady state and in the
thermodynamic limit $(L \to \infty)$ it must diverge as the
correlation time as $p$ approaches $p_c$. For a finite-sized
system (substrate) at $p = p_c$, $w(L, \infty) \propto
L^{\nu_\parallel / \nu_\perp}$ which gives the roughness exponent :

\begin{equation}
\alpha _{p = p_c} = {\nu_\parallel \over \nu_\perp} \, .
 \label{eq:alpha-A}
\end{equation}

\noindent Similarly the crossover time $t_{\times}$ is another way
of defining the characteristic timescale for the model and shall
have the form $t_{\times} \propto L^{\nu_\parallel / \nu_\perp}$ at $p = p_c$.
Therefore the dynamic exponent is given by :

\begin{equation}
z = {\nu_\parallel \over \nu_\perp} \, .
 \label{eq:z-A}
\end{equation}

\noindent Since the phase transition of PCA is believed to be in the
universality class of Directed Percolation~\cite{Bhattacharyya1998},

\begin{equation}
\alpha _{p = p_c} = z = {\nu^{DP}_\parallel \over \nu^{DP}_\perp}
 \approx 1.58 \, .
 \label{eq:DPvalue}
\end{equation}

\noindent Therefore the growth exponent at $p = p_c$ becomes equal
to unity :

\begin{equation}
\lambda _{p = p_c} = {\alpha _{p = p_c} \over z} = 1 \, .
 \label{eq:lambda-A}
\end{equation}

\indent To confirm the results obtained above from scaling considerations
Monte Carlo simulations of the growth process in Model A were
performed on substrates of sizes $L = 100$ to $L = 10000$.
A disordered substrate is constructed by assigning to each
lattice site the value $1$ with probability $\frac{1}{2}$ and
$0$ otherwise. For each substrate size $2000$ independent
simulations were done for the purpose of averaging out
statistical fluctuations.

\indent Figure 3 shows the results of applying scaling relation
(\ref{eq:scale2}) to the simulation data for the surface width.
Data for $w(L, \infty)$ {\em vs} $L$ in the pinned phase, for
different values of $p$ close to $p_c$, collapse to a single
curve for the following values of the transition point and
the critical exponents :

\begin{equation}
p_c = 0.7513 \pm 0.0004 \, ;        \label{eq:simul1a}
\end{equation}

\begin{equation}
\nu_\perp = 1.1 \pm 0.01 \, ,~~~~~~~~~\nu_\parallel = 1.73 \pm 0.01 \, .
 \label{eq:simul1b}
\end{equation}

\noindent Similarly for the moving phase, values of the critical point
and exponents were obtained by collapsing the simulation data for
$w(L, \infty)$ {\em vs} $L$ for different values of $p$ close to $p_c$ :

\begin{equation}
p_c = 0.7511 \pm 0.0004 \, ;        \label{eq:simul2a}
\end{equation}

\begin{equation}
\nu_\perp = 1.1 \pm 0.01 \, ,~~~~~~~~~\nu_\parallel = 1.74 \pm 0.01 \, ,
 \label{eq:simul2b}
\end{equation}

\noindent These are found to agree, within the limits of error,
with the corresponding values for the pinned phase. Error bars
indicated here and elsewhere in the paper are statistical errors.

\indent The estimate of the depinning transition point $p_c$
obtained here agrees well with the value of the critical point
of the PCA measured by using defect dynamics~\cite{Bhattacharyya1998}.
The estimates of the critical exponents are in good agreement
with the values for the universality class of Directed Percolation
~\cite{Jensen1996}.

\indent To measure the surface exponents for $p = p_c$ simulations were
performed at $p = 0.7512$. The time taken for the PCA to reach a
steady state is very large and the width of the surface could not be
measured till saturation. Therefore a direct determination of
$\alpha$ and $z$ at $p_c$ was not possible with the
available computer facility. However the surface width (averaged
over $5000$ intial configurations) was observed to grow linearly
on a substrate of size $L= 10^4$ for $10^5$ times-steps (Figure 4),
thus establishing $\lambda_{p = p_c} = 1$ and $\alpha _{p = p_c} = z$,
in agreement with equation~(\ref{eq:lambda-A}).
Analyses of simulation data far from $p_c$ show that the
\lq rough\rq~ regime occurs for only very small system sizes
($L \ll L_\times$) and the dependence of $w(L, \infty)$ on $L$
is logarithmic for the most part of the crossover region.
The \lq rough\rq~ regime grows as $p$ approaches $p_c$ and $w(L, \infty)$
becomes entirely a power-function of $L$ at $p = p_c$.
Thus a true \lq rough\rq~surface (in the sense of $\alpha > 0$)
exists only at the transition point.

\indent The last result of this section is concerned with the
velocity of the surface in the moving phase. Simulations of the
growth process $A$ for different values of $p$ show that the
surface in the moving phase attains the maximum velocity
asymptotically, $v_\infty = 1$, irrespective of the value of $p$
(Figure 5). Consequently $v_\infty$ cannot serve as the
order parameter for the depinning transition in Model $A$. The
appropriate order parameter is the asymptotic density of columns
with the maximum height which is equal to the density of
occupied sites in the steady state of the generating PCA
and hence it decays by the power-law (\ref{eq:PCA-order})
with the critical exponent $\beta = \beta^{DP} \approx 0.276$.

\section{Properties of Model $B$}

\indent Model $B$ defines a deposition process of the
solid-on-solid type, i.e., the bulk of the aggregate is
compact (no vacancies) and there are no surface overhangs.
Consider first the moving phase $(p > p_c)$.
Since there is no interaction between the columns of the
aggregate, correlations do not develop along its surface;
hence the surface width in the moving phase does not saturate.
When the PCA evolves from a disordered initial state the
process can be described by a continuum equation similar to
the one describing a ordinary random deposition
process~\cite{Barabasi1995} :

\begin{equation}
{\partial \over {\partial t}} h(x, t) = F + \eta(x, t),
 \label{eq:continuum}
\end{equation}

\noindent where $F$ is the average number of particles arriving
at a site and $\eta$ is the noise term (with zero configurational
average : $\langle \eta(x, t) \rangle = 0$) that describes
the fluctuations in the deposition process. The difference
between the random deposition model and this model lies in the
noise correlations. While the noise in random deposition is
uncorrelated, the spatial and temporal correlations developing
in the generating PCA appear as noise correlations in model $B$.
Since the initial state of the PCA is disordered the growth
process in the first time-step looks like random deposition.
But correlations in space and time develope in the PCA
configuration as it evolves owing to the nearest-neighbour
interactions defined in the evolution rule (\ref{eq:model}).
For values of $p$ away from $p_c$, the correlation length
$\xi$ and the correlation time $\tau$ of the PCA are finite
which means the noise in the deposition process is correlated
over a short range described by exponential decay of the
correlation function~\cite{Barabasi1995} :

\begin{equation}
\langle \eta (x, t) \eta (x', t') \rangle \sim
e^{-|x - x'|/\xi} \: e^{-|t - t'|/\tau}.    \label{eq:short-noise}
\end{equation}

\noindent The effects of the correlated noise is prominent for
times less than $\tau$. Since deposition at time-step $t + 1$
is allowed only at or next to sites where deposition has taken
place in the previous time-step $t$ [along with the restrictions
set by the PCA evolution rule (\ref{eq:model})] the height
fluctuations about the mean value are larger and hence the growth
of the surface width is faster than that for random deposition.
For times greater than $\tau$ the noise appears uncorrelated
and the growth exponent $\lambda$ will reduce to the random
deposition value $\lambda^{RD} = \frac{1}{2}$. As $p$
approaches $p_c$, $\xi$ and $\tau$ increase and it takes longer
for the growth process to reach the random deposition limit.
This is illustrated in Figure 6.  Finally, at $p = p_c$,
both $\xi$ and $\tau$ diverge and the correlations in the
noise are long-ranged, represented by a power-law decay of
the noise correlation function~\cite{Barabasi1995} :

\begin{equation}
\langle \eta (x, t) \eta (x', t') \rangle \sim
|x - x'|^{- 2 \beta / \nu_\perp} \: |t - t'|^{- 2 \beta / \nu_\parallel}.
 \label{eq:long-noise}
\end{equation}

\noindent where $\beta$, $\nu_\perp$ and $\nu_\parallel$ are the
critical exponents for the order parameter, correlation length
and correlation time of the PCA respectively. The asymptotic value
of the growth exponent $\lambda$ at $p_c$ can be derived from
the continuum equation (\ref{eq:continuum}) where $\eta(x, t)$
follows the correlation function (\ref{eq:long-noise}) instead
of the delta-correlated noise of a random deposition process. 
The surface width for a infinite substrate is found to
increase as the following power function of the time :

\begin{equation}
w(\infty, t) \propto t^{1 - \beta / \nu_\parallel}.    \label{eq:grow-B}
\end{equation}

\noindent Using the indication that the phase transition in the
PCA belongs to the universality class of Directed Percolation
\cite{Bhattacharyya1998}, the value of the growth exponent at $p_c$
is expected to be :

\begin{equation}
\lambda _{p = p_c} = 1 - {\beta^{DP} \over \nu_\parallel^{DP}}
 \approx 0.841.    \label{eq:lambda-B}
\end{equation}

\noindent At this point the surface grown on a finite substrate
enters the pinned phase.

\indent The behaviour of the velocity of the surface in the moving phase
can be analysed as follows. The average height of the surface above
the substrate at time-step $t$ is equal to the number of particles
deposited per site in the interval $[0, t]$ averaged over the
entire system. The velocity of propagation $v_t$ of the surface at
time-step $t$, defined by Eq. (\ref{eq:velocity}), is therefore equal
to the average number of particles deposited per site in the $t-$th
time-step. This number is precisely the density of occupied sites
$\rho_t$ of the generating PCA at time $t$ :

\begin{equation}
v_t = \rho_t.
\end{equation}

\noindent In the steady state, as $p$ appoaches $p_c$,
the asymptotic velocity $v_\infty$ of the surface will thus decrease
to zero by the power-law followed by $\rho_\infty$. This implies,
after comparing equations (\ref{eq:PCA-order}) and (\ref{eq:grow-order}),
that $\theta = \beta$. Since the phase transition of the PCA is known
to be in the universality class of Directed Percolation
\cite{Bhattacharyya1998}, the critical velocity exponent is given by :

\begin{equation}
\theta = \beta^{DP} \approx 0.276.    \label{eq:theta-B}
\end{equation}

\indent The model $B$ in the moving phase was studied numerically
by Monte Carlo simulations of the growth process on a one-dimensional
lattice of $L = 10^4$ sites for various values of $p$ ($p > p_c$)
and results for each value of $p$ were averaged over $10^3$ independent
realisations of the process obtained by constructing the disordered
initial states of the PCA independently of one another.
Figure 6(a) shows the surface width as a function of time
for different values of $p$, drawn on double-logarithmic scale.
To determine the growth exponent $\lambda$ an effective
exponent $\lambda_t$ is defined as the local slope
of the curves drawn in Figure 6(a) :

\begin{equation}
\lambda_t = {\log [w(L, t) / w(L, t/b)] \over \log b}.  \label{eq:lambda-eff}
\end{equation}

\noindent With $b = 5$, $\lambda_t$ was calculated for different
values of $p$ upto $t = 10^5$ time-steps and plotted against $1/t$
(Figure 6b). All curves for $p > p_c$ tend to the random
deposition value $\lambda^{RD} = \frac{1}{2}$; though the curves
for $p$ close to $p_c$ do not reach this value in $10^5$
time-steps they clearly show a tendency to do so at a larger $t$.
For $p = 0.7512 (\approx p_c)$, in order to remove finite-size
effects, the growth exponent was obtained by extrapolating the
maximum values of the effective exponent measured for different
substrate sizes :

\begin{equation}
\lambda _{p = p_c} \approx \max \, [\lambda_t (p = 0.7512)]
 = 0.837 \pm 0.011,    \label{eq:sim3a}
\end{equation}

\noindent which is close to the expected value for an infinite
substrate [equation (\ref{eq:lambda-B})]. In the curve for
$p = 0.7512$ in Figure 6(b) a decrease observed in the value
of $\lambda_t$ after it reaches a maximum is due to the finite
size of the substrate used in the simulation and it shows the
beginning of a crossover to saturation of the surface width,
the system having just entered the pinned phase at $p_c$.

\indent The asymptotic velocity $v_\infty$ of the surface obtained
from the simulation data (Figure 7) show that the power-law behaviour
of $v_\infty$ as $p \to p_c^+$ [equation(\ref{eq:grow-order})]
is best satisfied with :

\begin{equation}
p_c = 0.7514 \pm 0.0004.    \label{eq:simul3b}
\end{equation}

\noindent This independent estimate of $p_c$ agrees well previous
estimates~\cite{Bhattacharyya1998} and those obtained from the
study of model $A$ in this paper. The value of the velocity
exponent was measured to be :

\begin{equation}
\theta = 0.274 \pm 0.007,    \label{eq:simul3c}
\end{equation}

\noindent which agrees, within the limits of error, with the expected
value given in equation (\ref{eq:theta-B}).
  
\indent Consider now the pinned phase. For $p < p_c$ growth
of the aggregate stops in finite time (which depends on the
parameter $p$) due to the appearence of the absorbing state
of the generating PCA. The time for which growth takes place
is equal to the {\em transient time} $T$ of the generating PCA,
i.e., the number of time-steps required by the PCA to evolve
from a disordered initial configuration to the absorbing zero
configuration. Like that of model $A$, the surface width $w(L, t)$
saturates after a crossover time $t_\times$ for a substrate
of finite size. The behaviour of the saturated width $w(L, \infty)$
too resembles that of model $A$ : after an initial $L$-dependent
regime it passes over to a $L$-independent regime (Figure 2c) beyond
a crossover length $L_\times$. Thus for $p < p_c$ the surface
in model $B$ is smooth (in the sense of $\alpha = 0$) in the
thermodynamic limit ($L \to \infty$). Only at the depinning
transition point $p_c$, where $L_\times$ diverges, the surface
is rough ($\alpha > 0$) for all substrate sizes and the Family-
Vicsek law (\ref{eq:scale3}) is expected to hold true.
Since the saturation of the surface width is forced by the PCA,
the crossover time $t_\times$ will be the same as that of model
$A$ which means the dynamic exponent at $p_c$ is given by the
Directed Percolation value :

\begin{equation}
t_\times (p = p_c) \propto L^z,~~~~~~~~~z = z^{DP} \approx 1.58. 
 \label{eq:z-B}
\end{equation}

\indent At $p_c$, just before the surface enters the moving
phase, the roughness exponent is given by :

\begin{equation}
\alpha _{p = p_c} = {z \cdot \lambda_{p = p_c}}
 = z^{DP} (1 - \beta^{DP} / \nu_\parallel^{DP}) \approx 1.33.
 \label{eq:alpha-B}
\end{equation}

\noindent This result for $\alpha$ could not be verified by
Monte Carlo simulations on available computer facility as it
takes enormous times for the surface width on substrates of
reasonable sizes to reach saturation at $p_c$.

\indent Since the surface width does not saturate in the moving
phase, it becomes infinitely large with time. Therefore the
depinning transition point also marks a kinetic roughening transition
from a smooth surface in the pinned phase to a infinitely rough
surface in the moving phase.

\section{Discussion}

\indent In the two models studied in this paper the
dynamics of surface growth are derived from the evolution
rule of a one-dimensional probabilistic cellular automaton.
While model $A$ presents the growth of an aggregate of
interacting particles, model $B$ presents the growth of
an aggregate of non-interacting particles deposited by
a correlated mechanism. This results in the
formation of a correlated surface in model $A$ and an
uncorrelated surface in model $B$. The correlated mechanics
of deposition in model $B$ leads to growth of the surface
width of the aggregate at the depinning transition point $p_c$
with an exponent $\lambda$ larger than that offered the
ordinary random deposition model. This result follows the
trend that the growth exponent increases in the presence of
correlations in the deposition process, observed previously
in ballistic deposition models~\cite{Meakin1989-90}.
Also the surface properties of the models at the depinning
transition are all related to the critical properties
of the generating PCA. It is yet to be seen the effect
of introducing in these models the additional dynamics
of surface restructuring.

\indent Finally, a comment on the scaling relation~\cite{Barabasi1995} :

\begin{equation}
\theta = (z - \alpha) \nu ,    \label{eq:scale4}
\end{equation}

\noindent which relates the exponents for roughness, growth and
velocity of the surface to the exponent $\nu$ for the correlation
length along the surface at a depinning transition. For model $A$
at $p_c$, $\nu = \nu_\perp$ and $\alpha = z$ so that $\theta = 0$;
this is true as the velocity in the moving phase is independent
of the level of noise $p$ which provides the driving force.
For model $B$ at $p_c$, the values of $\theta$, $z$ and $\alpha$
from equations (\ref{eq:theta-B}), (\ref{eq:z-B}) and
(\ref{eq:alpha-B}) give $\nu = \nu_\perp$; this is not true.
Since there is no inter-columnar interaction at the
surface of the aggregate, correlations do not develop along the
surface and an exponent $\nu$ cannot be associated with it; the
scaling relation (\ref{eq:scale4}) therefore does not hold in this
case.

\section*{Acknowledgement}

\indent I am grateful to Bikas K. Chakrabarti for
his critical comments on the work. A part of this work was
supported by CSIR, Government of India.

\newpage

\section*{Figure Captions}

\begin{description}

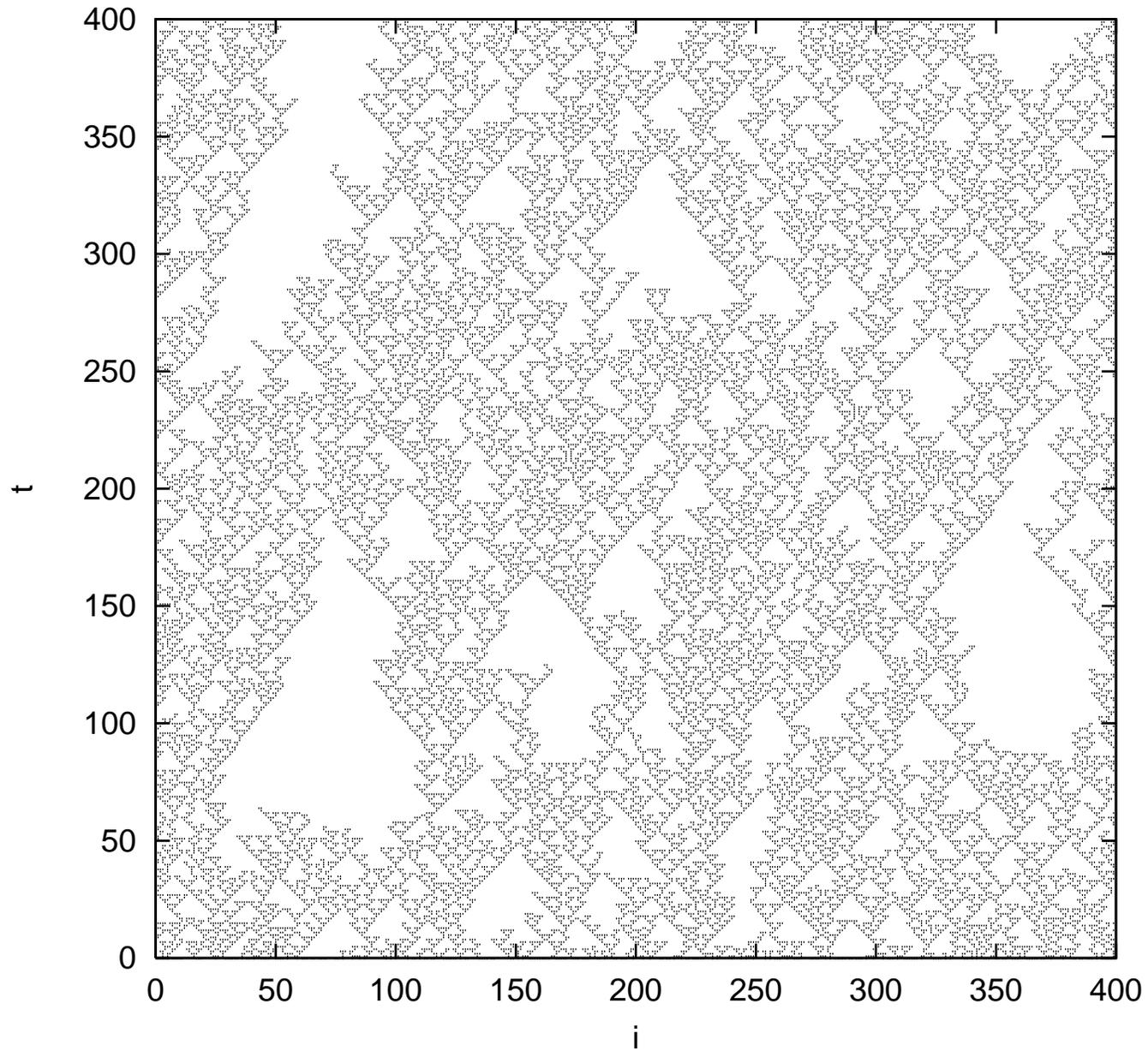
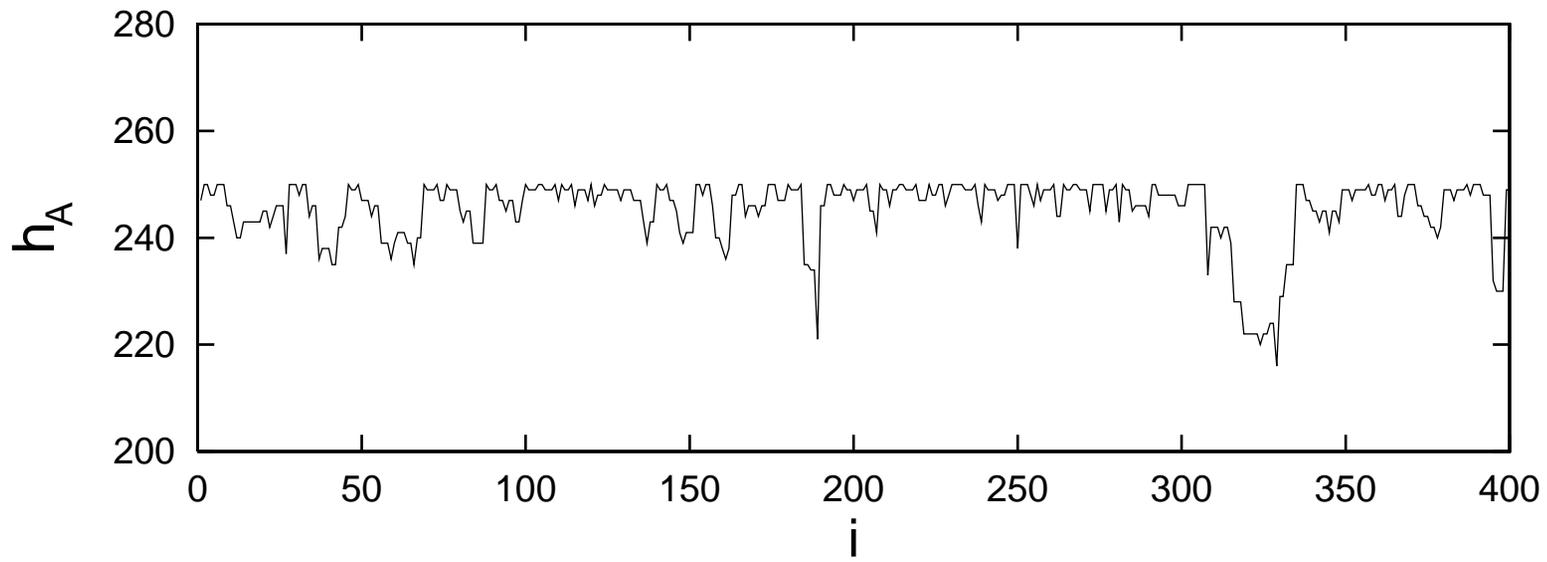
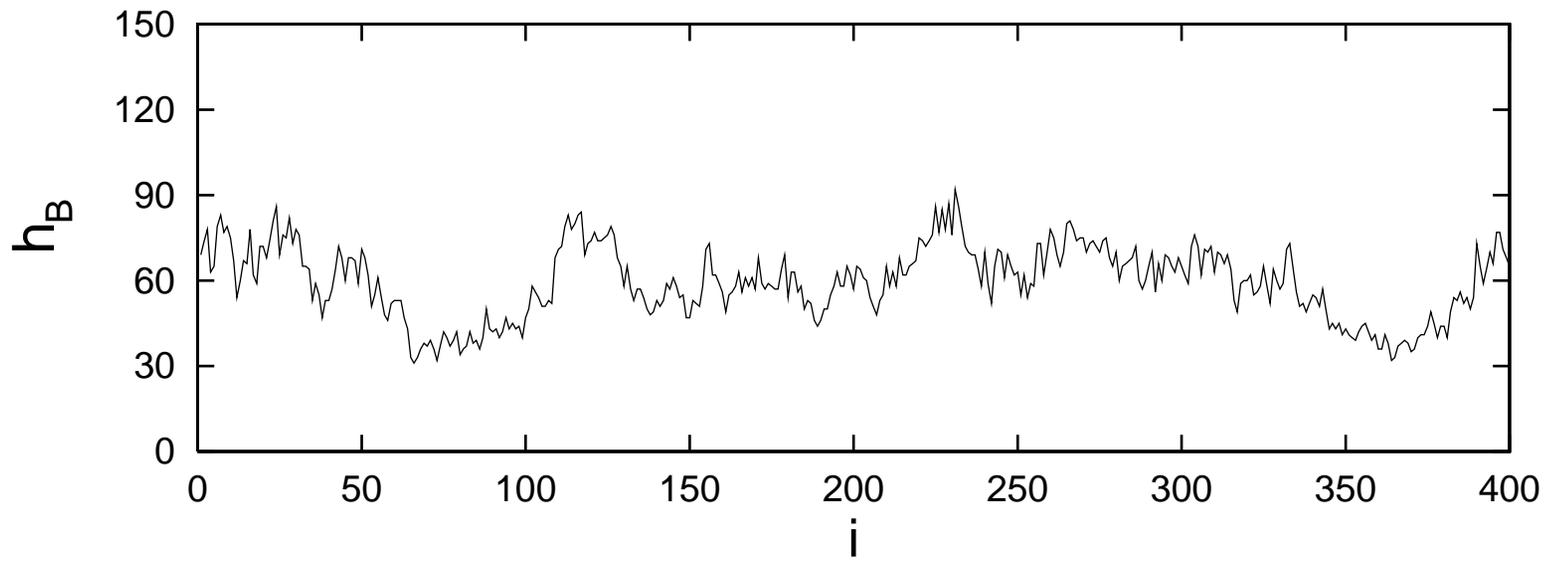
\item [{Figure 1.}] (a) An example of the evolution of the
probabilistic cellular automaton following the rule in equation (1)
for $p = 0.8$. Sites $(i, t)$ on the space-time plane with
automaton value $a_i^{(t)} = 1$ are marked dark while those
with automaton value $a_i^{(t)} = 0$ are left blank.
(b) The surface of model $A$ and (c) that of model $B$
at the 250-th time-step of the automaton evolution shown in (a).
(d) Schematic diagram showing the construction of the surfaces
of models $A$ and $B$ : model $A$ is constructed from the PCA pattern
by filling all vacant sites below the highest occupied site in
every column whereas model $B$ is constructed by considering the
occupied sites in the PCA pattern as particles and dropping them
one above the other in a ordered way in every column. The surfaces
are shown by lines joining the highest point in each column.

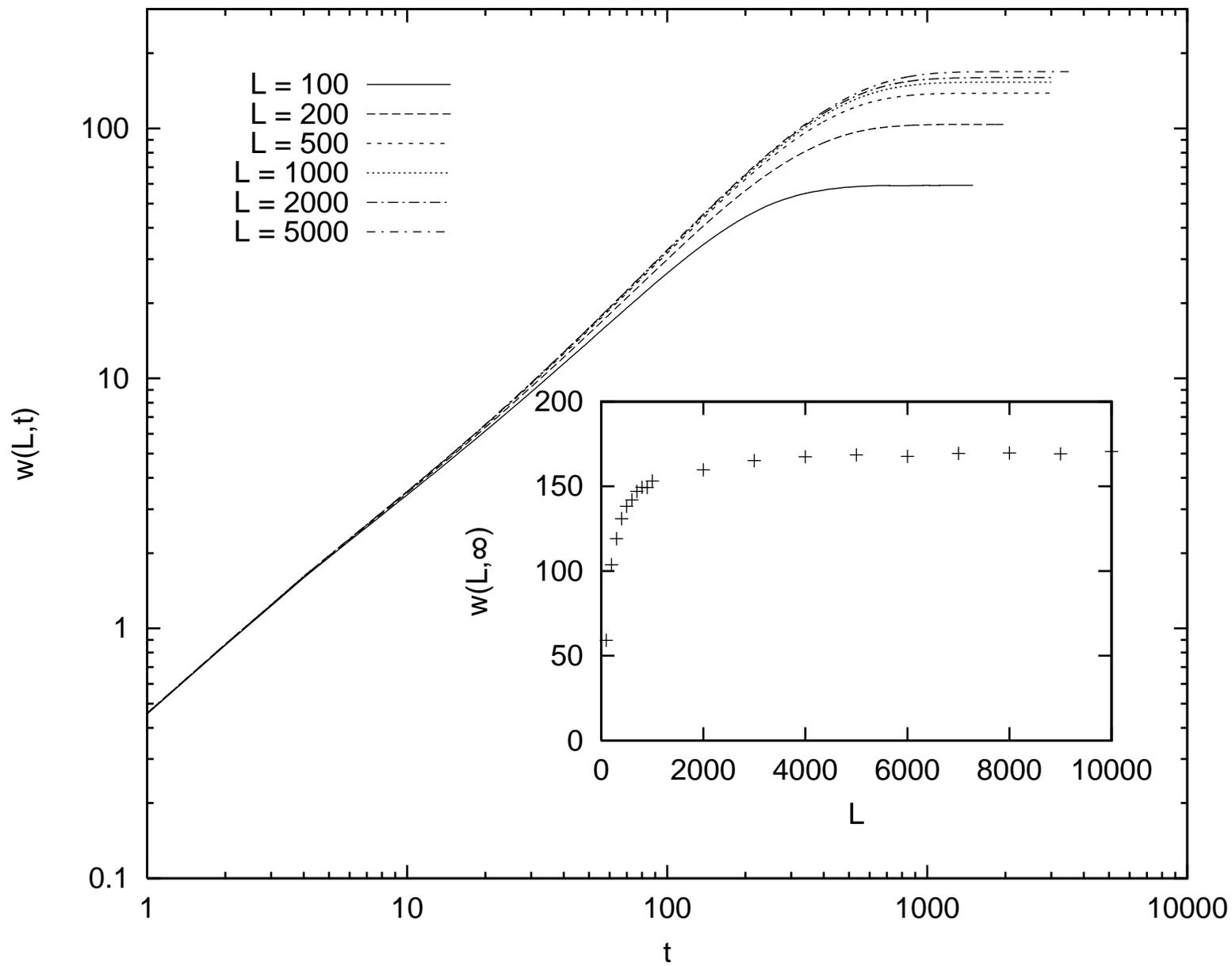
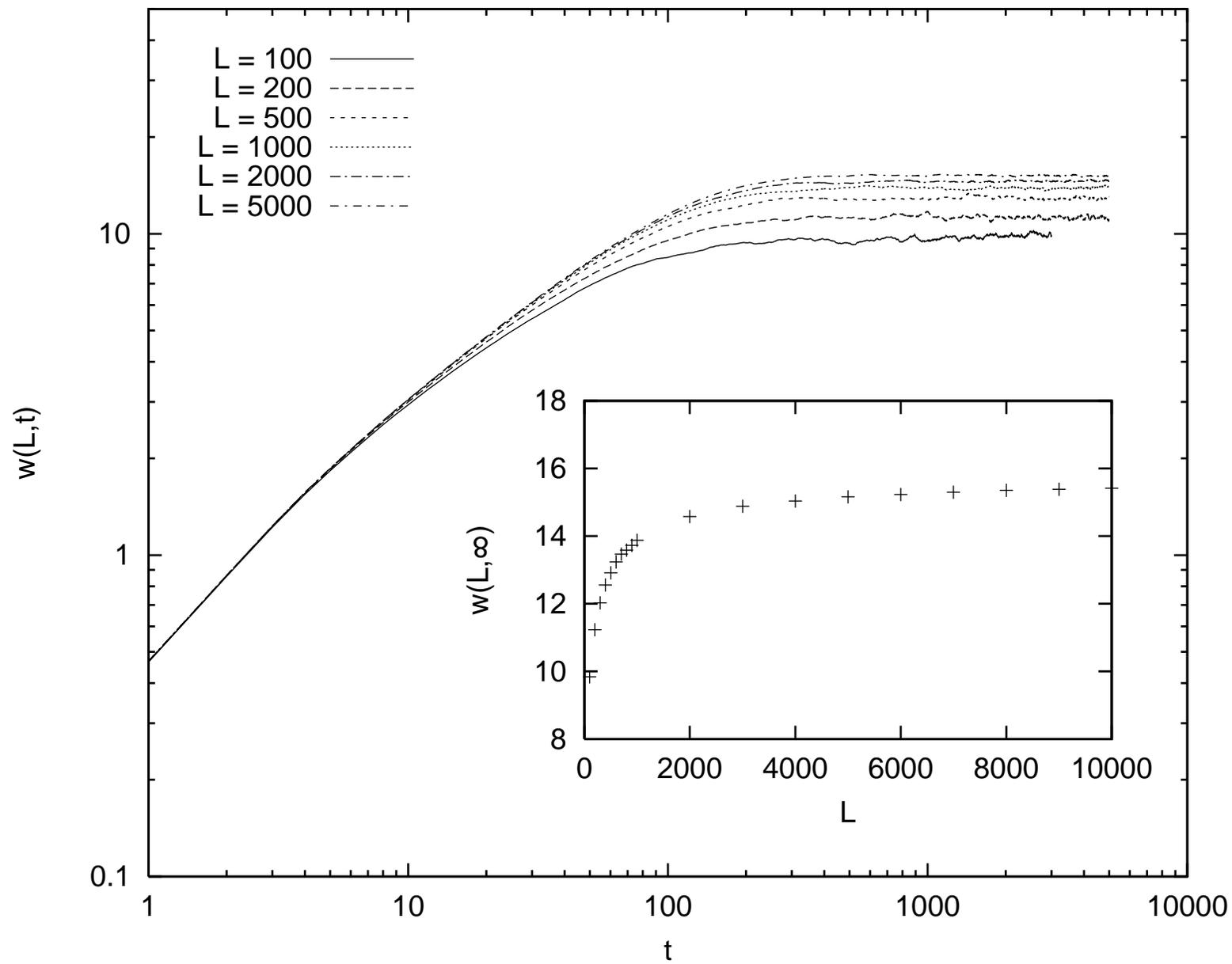
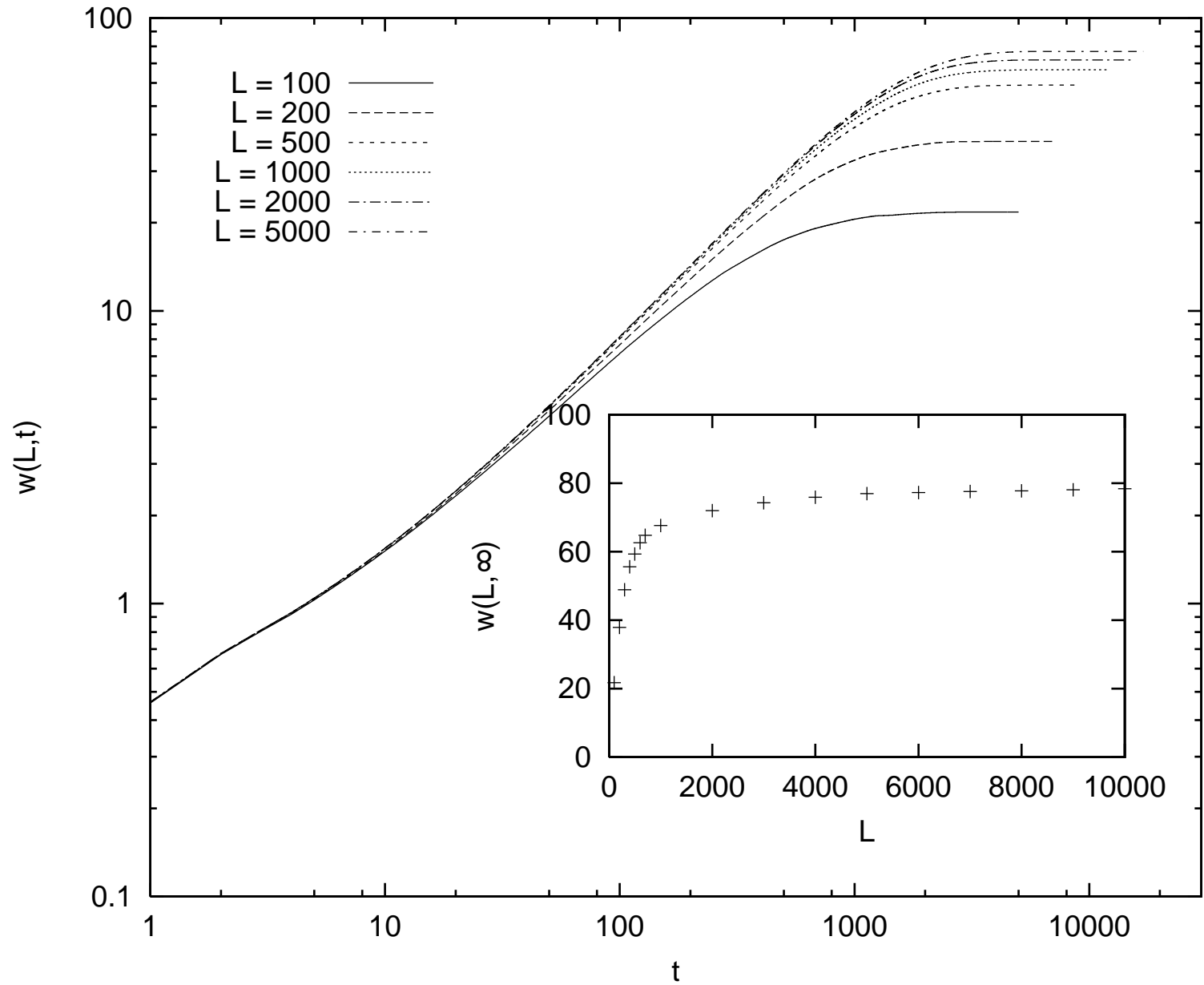
\item [{Figure 2.}] The development of the surface width in time on
substrates of different sizes : (a) model $A$ in the pinned
phase ($p = 0.7$), (b) model $A$ in the moving phase ($p = 0.8$),
(c) model $B$ in the pinned phase ($p = 0.73$).

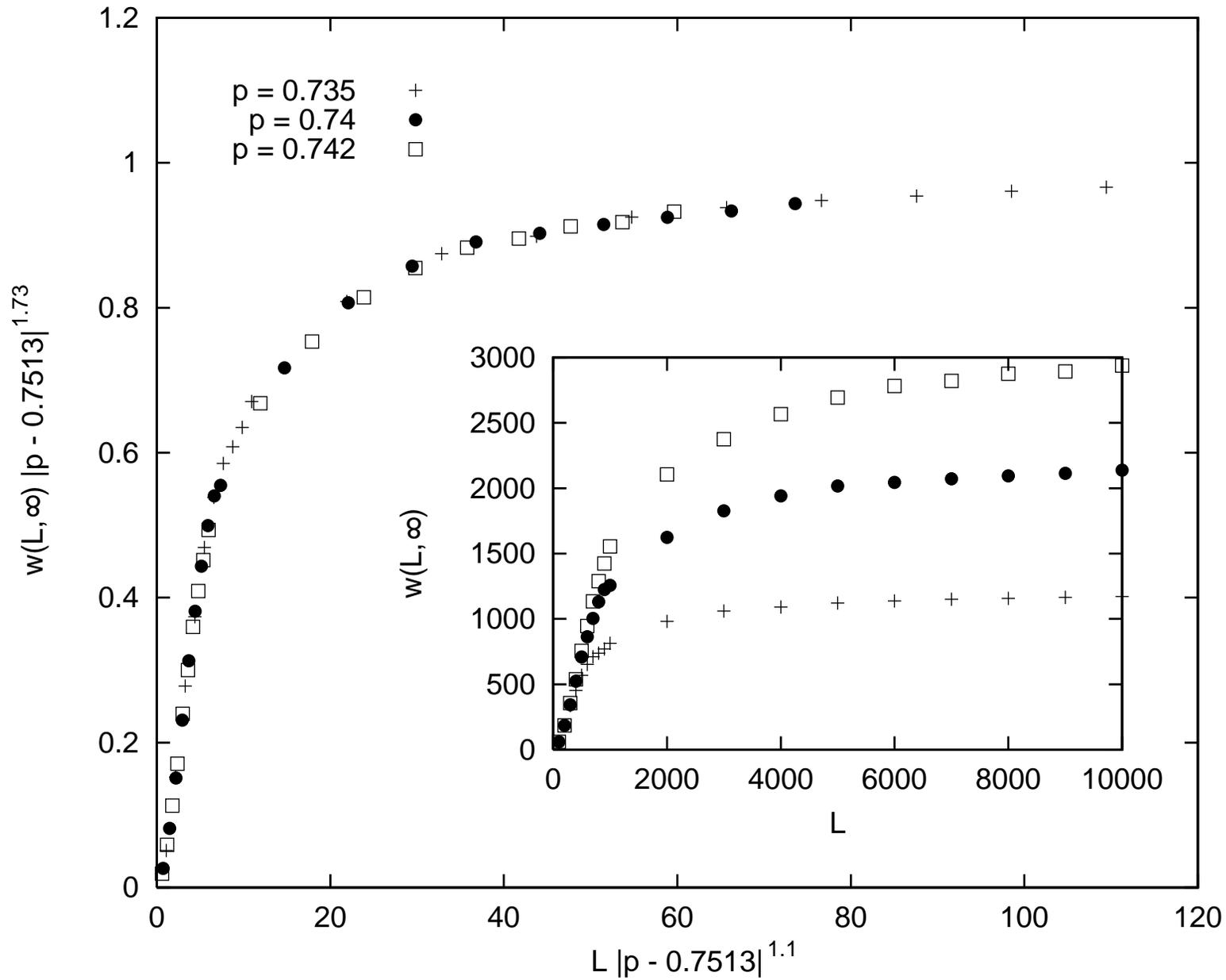
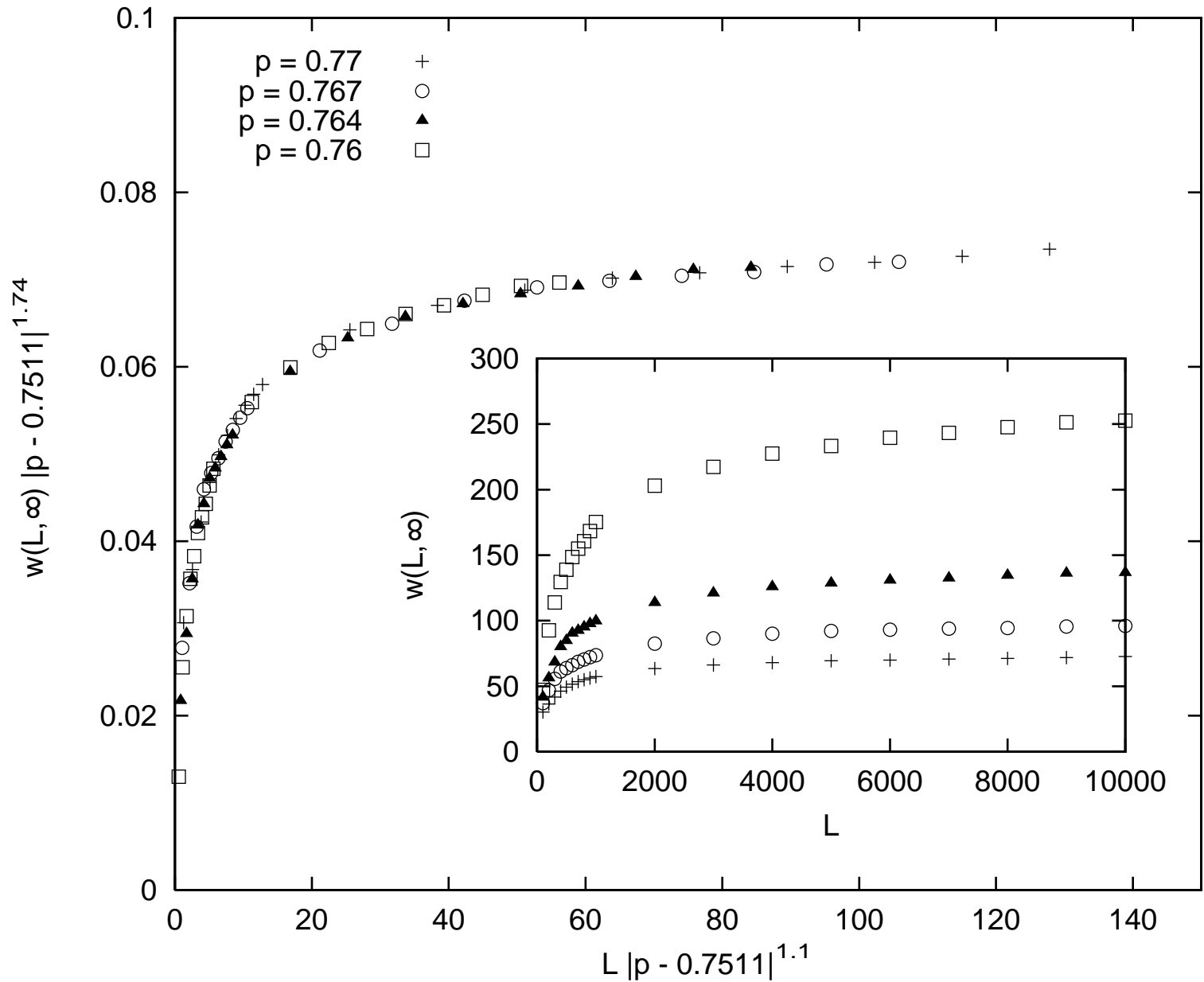
\item [{Figure 3.}] Data collapse obtained by applying
the finite-size scaling relation in equation (10) to the
simulation data for the saturated surface width of model $A$
in the (a) pinned and (b) moving phases.

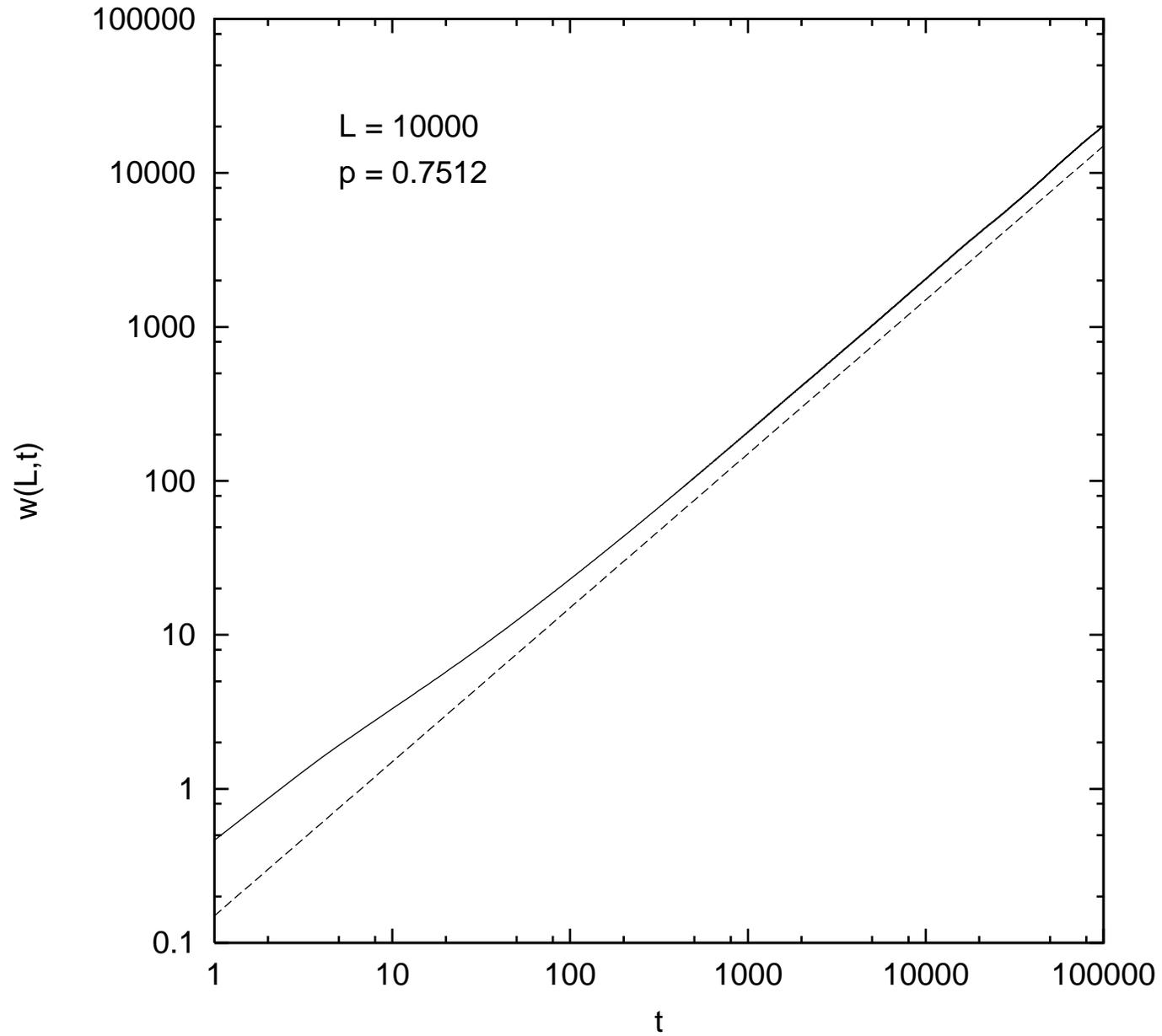
\item [{Figure 4.}] The growth of the surface width of
model $A$ at the depinning transition point $p_c \approx 0.7512$.
The solid line shows the result of simulation on a lattice
of $10^4$ sites; the dashed line represents a linear growth
of the surface width : $w(L, t) \propto t$. At large times
the surface width is seen to grow linearly with time as it
becomes parallel to the dashed line.

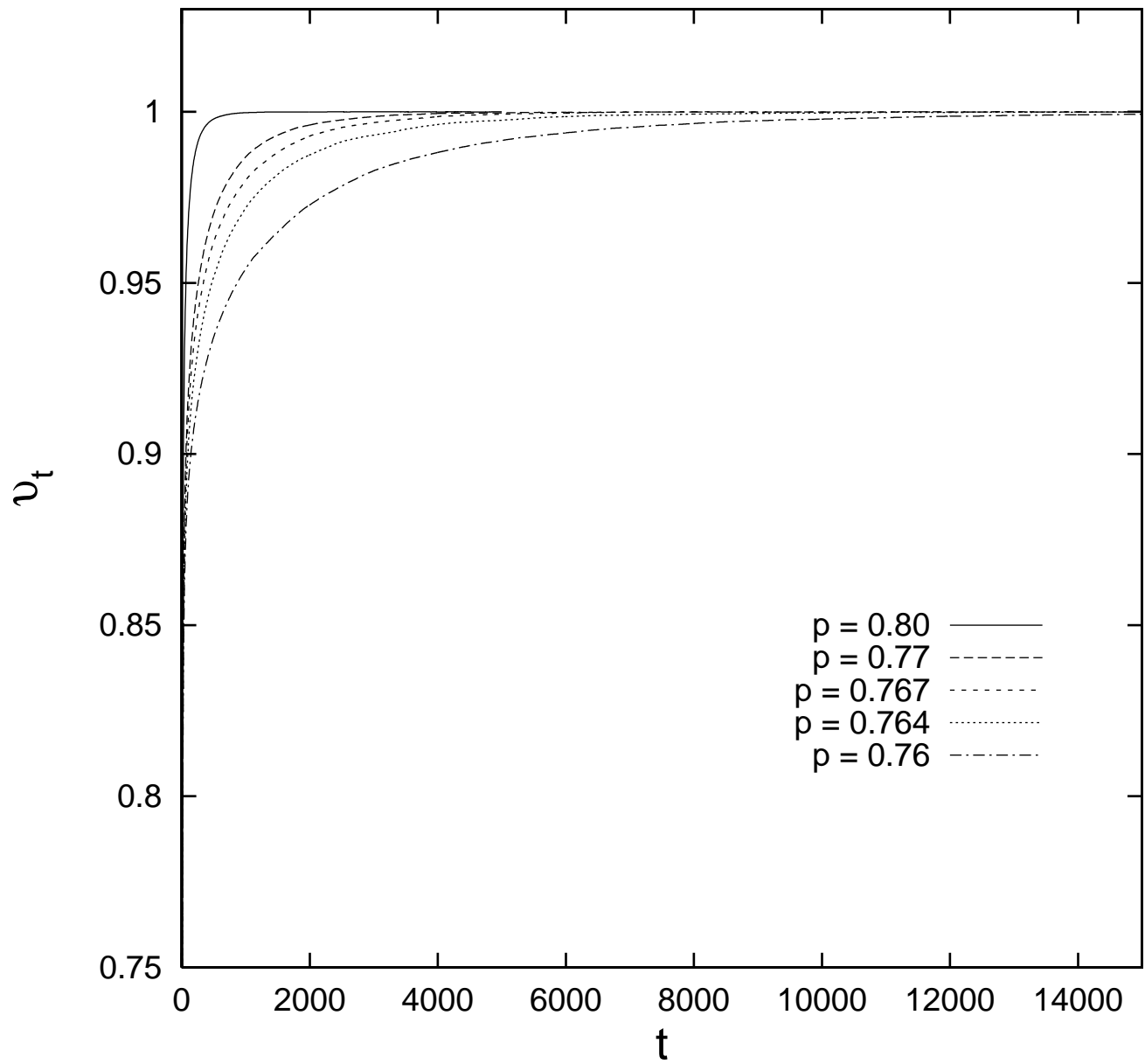
\item [{Figure 5.}] The average velocity of the surface,
defined by equation (4), in the moving phase of model $A$
is plotted against time. For all values of $p$, a few of
which are shown here, the velocity asymptotically attains
the maximum possible value of one lattice constant per
time-step.

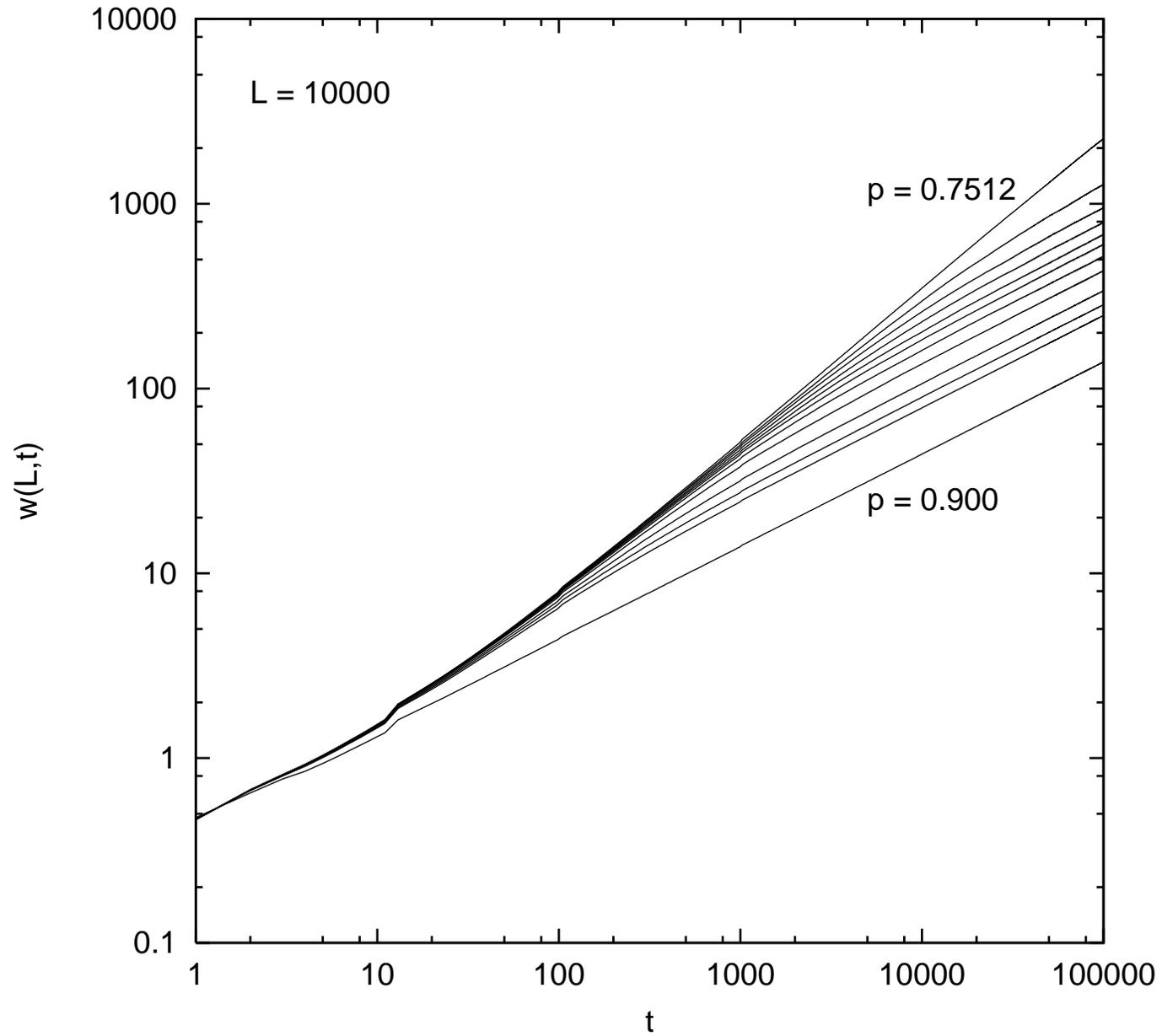
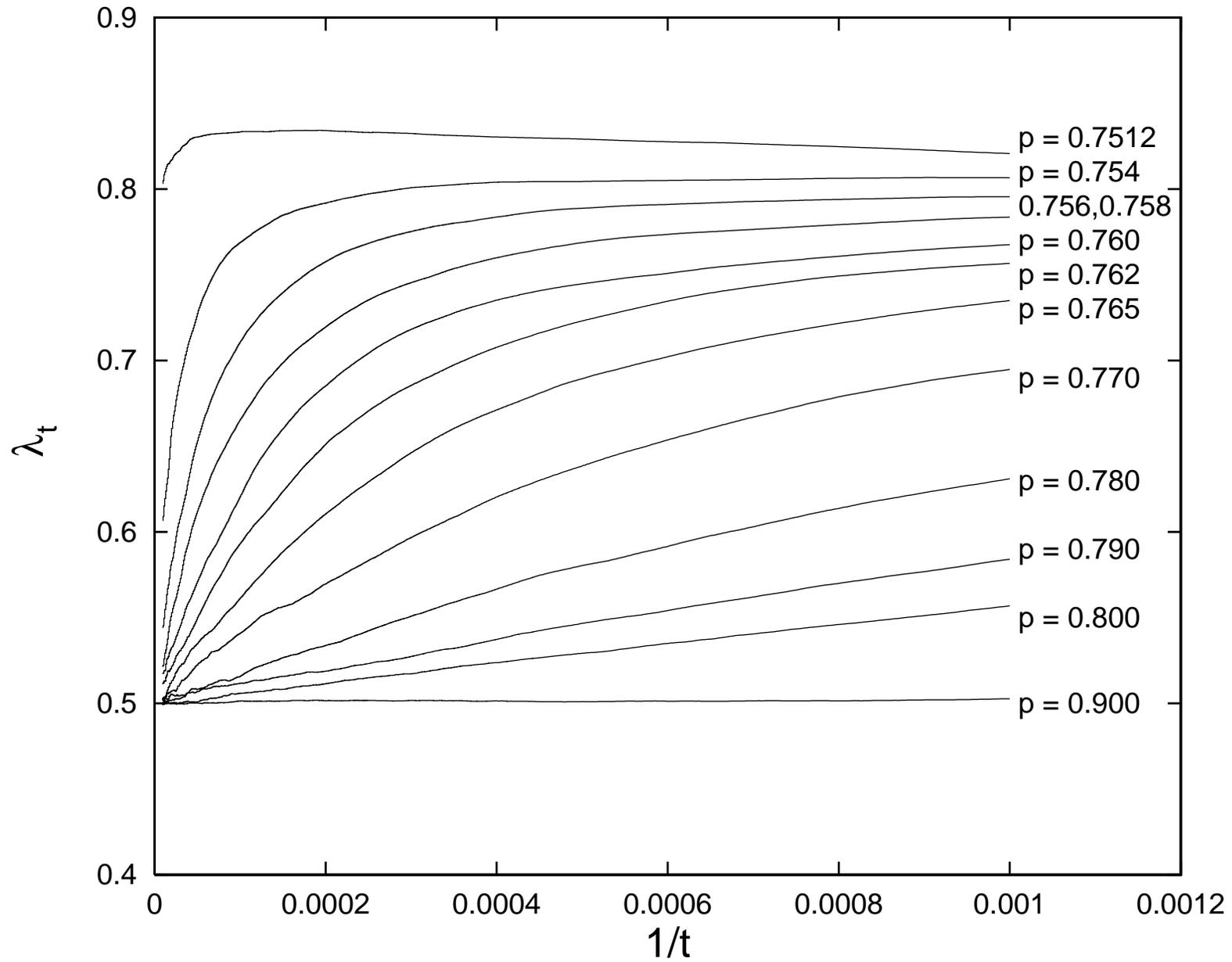
\item [{Figure 6.}] (a) The growth of the surface width
of model $B$ in time for different values of $p$ in the
moving phase. Simulation data are shown for the following
values of $p$ : 0.7512 (top), 0.754, 0.756, 0.758, 0.76,
0.77, 0.78, 0.79, 0.8 and 0.9 (bottom). The curves do not
appear to saturate even after $10^5$ time-steps of evolution,
which agrees with the theory. (b) The effective growth
exponents $\lambda_t$ derived as local slopes of the
curves shown in (a). The curves for $p = 0.754,~\ldots~, 0.9$
tend to the random deposition value $\lambda = 0.5$ as
$t \to \infty$. The downward bending of the curve for
$p = 0.7512$ is a signature of relaxation toward a
saturated state; this is not easily evident from the
corresponding curve in (a).

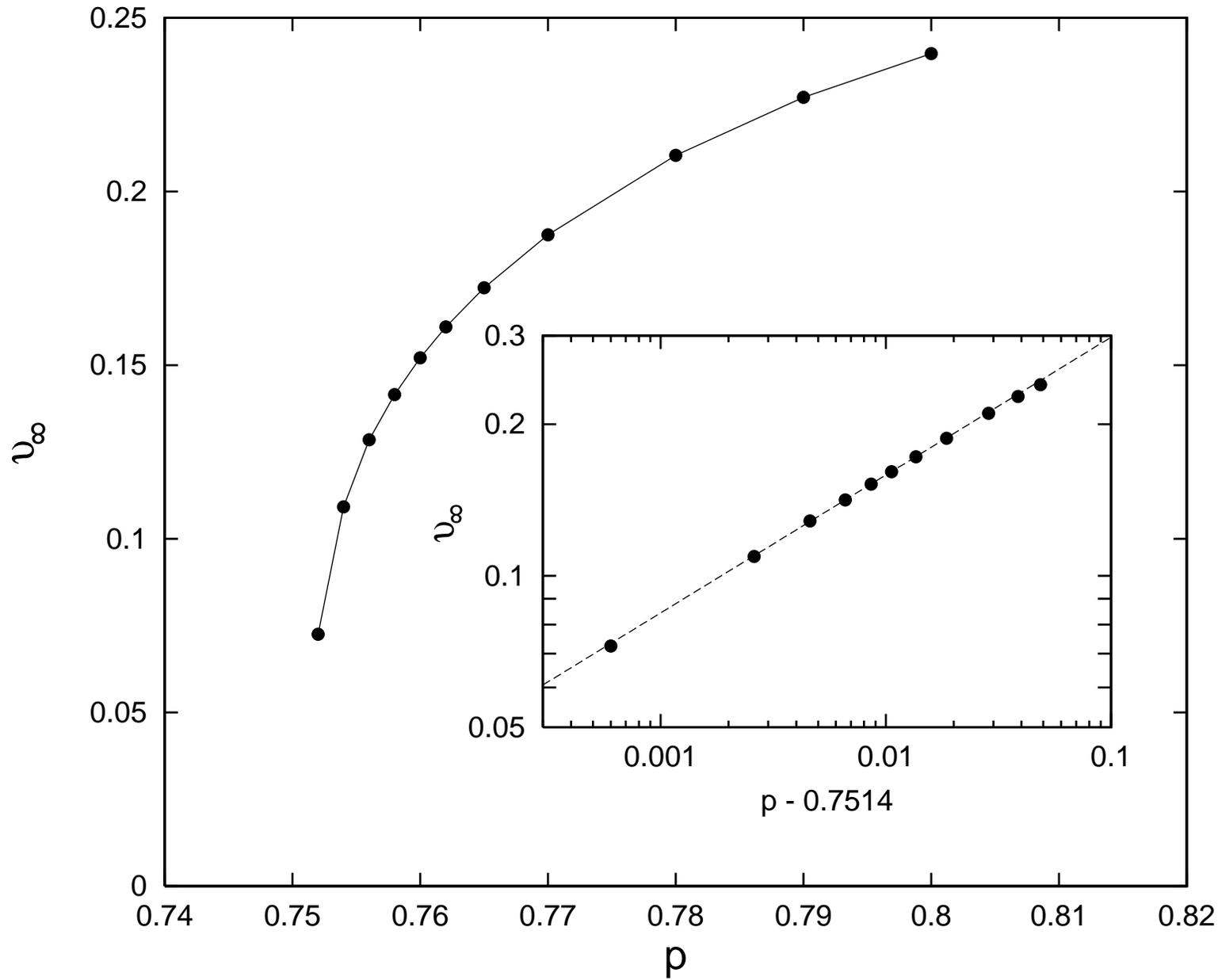
\item [{Figure 7.}] The asymptotic velocity of model $B$
in the moving phase for different values $p$ near $p_c$.
The inset shows that it decreases to zero by a power-law
as $p$ approaches $p_c$. The slope of the dashed line
gives the value of the velocity exponent $\theta$ (see text).

\end{description}

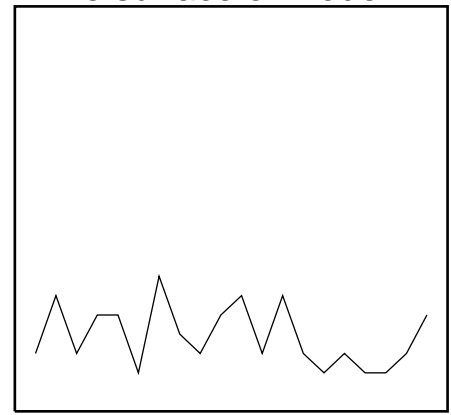
\end{document}